\documentclass[tcfd, numbers]{svjour}

\bibpunct{[}{]}{;}{a}{,}{,}

\usepackage{amsmath}
\usepackage{graphicx}
\usepackage{subfigure}
\usepackage{color}
\usepackage{epstopdf}

\begin{document}


\title{
Analytic models of heterogenous magnetic fields for liquid metal
flow simulations}

    \Author{E.~V.~Votyakov, S.~C.~Kassinos, X.~Albets-Chico}
    {Computational Science Laboratory UCY-CompSci, Department of
    Mechanical and Manufacturing Engineering, University of Cyprus, 75
    Kallipoleos, Nicosia 1678, Cyprus}



%
%

\commun{Communicated by }
%
\date{Received date and accepted date}
%

\abstract{
    A physically consistent approach is considered for defining an external magnetic field
    as needed in computational fluid dynamics problems involving magnetohydrodynamics (MHD).
    The approach results in simple analytical formulae that can be used in numerical studies where an
    inhomogeneous magnetic field influences a liquid metal flow. The resulting magnetic
    field is divergence and curl-free, and contains two components and parameters to vary.
    As an illustration, the following examples are considered: peakwise, stepwise,
    shelfwise inhomogeneous magnetic fields,  and the field induced by a
    solenoid. Finally, the impact of the streamwise magnetic field component is shown qualitatively
    to be significant for rapidly changing fields.
}

\titlerunning{Analytic models of heterogenous magnetic fields}
\authorrunning{E.~V.~Votyakov, S.~C.~Kassinos, X.~Albets-Chico}

\maketitle


There are few examples in recent history, when things being
obvious to particular specialists, remained unexploited by people
working in conjugated fields. These include, for instance, the
Fast Fourier Transform (FFT), which was originally used by Gauss
in 1805 and then several times rediscovered by Lanczos and
Danielson in 1942, and Cooley and Tukey in
mid-1960s\footnote{Press, W. H.; Flannery, B. P.; Teukolsky, S.
A.; and Vetterling, W. T. "Fast Fourier Transform." Ch. 12.2 in
Numerical Recipes in FORTRAN: The Art of Scientific Computing, 3rd
ed. Cambridge, England: Cambridge University Press, Page 498,
2007; see also
http://mathworld.wolfram.com/FastFourierTransform.html}. Another
and more specific example is the so-called Savitzky-Golay
smoothing filter. The least square method being the base of the
filter has been formulated hundred years before experimentalists
working with spectra started to use it to treat their data. The
paper that popularized this method for the experimentalists is one
of the most widely cited papers in the journal \textit{Analytical
Chemistry}\footnote{A. Savitzky and Marcel J.E. Golay (1964).
Smoothing and Differentiation of Data by Simplified Least Squares
Procedures. Analytical Chemistry, 36: 1627–-1639; see also
http://en.wikipedia.org/wiki/Savitzky-Golay\_smoothing\_filter}.

What is considered in this letter is not as far reaching as the
two aforementioned examples, nevertheless, we believe it will help
people studying numerically the flow of liquid metals under the
influence of an inhomogeneous magnetic field.
Also, this letter is complementary to the discussion about
magnetic field models given earlier in \cite{Votyakov:JFM:2007}.
The cited work dealt with a 3D distribution of heterogeneous
magnetic fields, which are of importance in the case of a magnetic
obstacle, while the current letter considers simpler 2D
distributions, which are needed, for example, for the proper
description of the flow through a fringing magnetic field.

Let us recall shortly the effects of magnetic fields on conducting
liquid flows. When the induced magnetic field is negligible, the
external magnetic field interacts with the moving liquid and
produces the Lorentz force which brakes the flow in the direction
of motion, see e.g. \cite{Davidson:book:2001}. This phenomenon is
heavily exploited in many practical applications
(\cite{Davidson:Review:1999}), such as electromagnetic stirring,
electromagnetic brakes, and non-contact flow measurements
(\cite{Thess:Votyakov:Kolesnikov:2006}).

In reality, one always deals with an inhomogeneous magnetic field
because creating a strong \textit{and} homogeneous magnetic field for
experimental needs remains a hard practical challenge.
Despite this fact, starting from the pioneering work of
\cite{Hartmann:1937} many theoreticians love to work mostly with a
homogeneous magnetic field. It is clear, that a constant magnetic
field is already responsible for main phenomena such as the
formation of the Hartmann and parallel layers. Nevertheless, a
constant field could not produce the well developed M-shaped
velocity profile frequently used for the electromagnetic brake.
Due to \cite{Kulikovskii:1968}, who showed that the flow under
the influence of a strong and slowly varying magnetic field can be subdivided in a
core and a boundary layer, people started to exploit theoretically
inhomogeneous magnetic fields. It was demonstrated that the flow
goes parallel to characteristic surfaces, but all the
calculations, including the numerical ones, were carried out
either by neglecting the second magnetic field component, see,
e.g. \cite{Sterl:1990}, \cite{Molokov:Reed:Fusion:2003},
\cite{Molokov:Reed:Magnetohydrodyn:2003}, \cite{Alboussiere:2004},
\cite{Kumamaru:etal:2004}, \cite{Kumamaru:etal:2007},
\cite{ni_current_2007} or by employing an especially curvilinear
channel \cite{Todd:1968} to match boundary conditions. In the case
of numerical simulations, there is no particular technical problem
preventing the inclusion of the second component of the
inhomogeneous magnetic field; nevertheless this has not been done.
A probable explanation for this is the absence of  suitable and
analytically simple models to define an inhomogeneous magnetic
field. Therefore, there is the need for convenient formulae that
could allow us to do so, and the goal of the paper is to provide a
simple method that can be used in numerical studies so that both
components can be varied in a consistent way. At the end of the
paper we discuss briefly the legitimacy of omitting the streamwise
component of the magnetic field, and show that in some cases this
might have been done due to large aspect ratios.

It is worth to note that there are examples where correct
expressions for a heterogenous magnetic field have been used for
MHD flow simulations. These are 2D numerical papers by
\cite{Cuevas:Smolentsev:Abdou:2006,
Cuevas:Smolentsev:Abdou:PRE:2006} who correctly applied formulae
taken from the book of \cite{McCaig:Book:1977}. Those 2D
simulations hired only the transverse component of the magnetic
field while the second component played no role. More
sophisticated 3D cases are given in \cite{Votyakov:PRL:2007,
Votyakov:JFM:2007}, where all three components of the magnetic
field are taken into account.

The necessity to have at least two nonzero components of the
inhomogeneous magnetic field (hereafter denoted as MF) follows
directly from the requirement that, in the flow region, an
externally applied field must be simultaneously divergence-free
$\nabla\cdot\bf{B}=0$ and curl-free $\nabla\times\bf{B}=0$. Thus,
if the transverse MF component varies along the streamwise
coordinate, the streamwise MF component must vary consistently
along the transverse coordinate. (The spanwise component can be
neglected without violation of the physical correctness, when the
magnetic field is two dimensional.) A vector field that is
simultaneously divergence-free and curl-free is known as a Laplace
vector field and it can be defined in terms of  the gradient of
any function $\eta$ which is harmonic in the flow region,
$\bf{B}=-\nabla\eta$, $\Delta\eta=0$. This function $\eta$ is
called the  magnetic scalar potential. Although this is the most
general approach, see e.g. \cite{McCaig:Book:1977}, it is not
quite convenient because boundary conditions for $\eta$ must be
defined for each specific case. However, what we need is a
magnetic field which either vanishes or goes monotonically onto a
constant level  far away from the central point where the
intensity of the field is maximal. So, among all the possible
harmonic functions, we may select those that do not vary at far
distances, and this forms the basis for the proposed methodology.

A simple, flexible, and physically consistent way to define an
inhomogeneous MF for parametric numerical needs is based on the
magnetic field induced by a single magnetic dipole. This field is
local in space and its magnetic scalar potential, which below is
called ''an elementary potential'', belongs to harmonic functions.
The whole magnetic field is the sum of local fields from the
single magnetic dipoles, therefore, the whole magnetic field can
also be represented as the gradient of a scalar function, i.e. as a
whole scalar potential. In other words, the single magnetic dipole
can be taken as an elementary unit in the appropriate spatial
distribution of magnetic sources. The field ${\bf B'}({\bf r,
r'})$ created at ${\bf r}=(x,y,z)$ by a single dipole ${\bf
m}=(0,0,m)$ located at ${\bf r'}=(x',y',z')$ is given by
\cite{Jackson:Book:1999}:
\begin{eqnarray}
    {\bf B'}({\bf r, r'})=\nabla\times\left(\nabla\frac{1}{|{\bf r-r'}|}\times{\bf m}\right)=
    ({\bf m}\cdot \nabla)\left(\nabla\frac{1}{|{\bf r}-{\bf
    r'}|}\right)=m\nabla\frac{\partial }{\partial z}\left( \frac{1}{|\bf{r}-\bf{r'}|}\right),
    \end{eqnarray}
where  we have used  few vector identities and omitted the
constant $\mu_0/(4\pi)$. Let the dipoles be distributed in the
finite region $\Omega$ outside of the flow. Then, it is easy to see
that the elementary $\eta'$ and the whole scalar potential $\eta$
are given by:
    \begin{eqnarray}
     \eta'(x,y,z,x',z',z')&=&-m\frac{\partial }{\partial z}\left( \frac{1}{|\bf{r}-\bf{r'}|}\right)
        =\frac{m(z-z')}{|\bf{r}-\bf{r'}|^3},\\
     \eta(x,y,z)&=&\int_\Omega  \eta'(x,y,z,x',z',z') dx' dy' dz'
     = \int_\Omega  \frac{m(z-z')}{|\bf{r}-\bf{r'}|^3} dx' dy' dz'\,.
    \end{eqnarray}
This is the simplest model of a permanent magnet occupying the
region $\Omega$ with the constant magnetic dipole distribution
$m$. By taking various $\Omega$, one can define any desired
inhomogeneous MF for parametric numerical needs and mimic roughly
various magnetic systems. Below in Fig.\ref{Fig:duct:MF} we give
few simple examples of two-dimensional, i.e. $B_y(x,y,z)=0$,
fields constructed in this way to represent typical magnetic field
configurations. Everywhere, $x$ (left to right) is streamwise, $y$
- spanwise (perpendicular to the plane of Fig.~\ref{Fig:duct:MF}),
and $z$ (down to up) is transverse coordinate.

Here it is worth noting that any two-dimensional  field can be
conveniently expressed through a complex representation.
Introducing the variable $\zeta=x+iz$, and its conjugate
$\zeta^{*}=x-iz$, one may define a complex function
     \begin{align}
            \beta(\zeta)=B_x+i\,B_z, \label{eq:beta}
     \end{align}
where $B_x=\rm{Re}(\beta)$, and $B_z=\rm{Im}(\beta)$. The function
$\beta(\zeta)$ is given below as well.

\textbf{Linear chain} of equal dipoles oriented parallel to
$z$ direction and infinitely extended in the $y$ direction at
$(x_0,z_0)$, $\Omega=\{x=x_0,-\infty\leq y \leq \infty, z=z_0\}$:
    \begin{align}
        \begin{split}
    \eta(x,y,z) &= \int_{-\infty}^{\infty} \int_{-\infty}^{\infty}
                    \int_{-\infty}^{\infty}
                 \frac{m(z-z')\delta(x'-x_0)\delta(z'-z_0)dx'dy'dz'}
                    {((x-x')^2+(y-y')^2+(z-z')^2)^{3/2}}
    \\
    &=\frac{2 m(z-z_0)}{(x-x_0)^2+(z-z_0)^2}=\frac{2 m \cos \varphi}{l} \label{eq:chainMF:eta}
        \end{split}
    \end{align}

    \begin{align}
    B_x(x,y,z)&=-\frac{\partial\eta(x,y,z)}{\partial x}=
        \frac{4 m(x-x_0)(z-z_0)}{\left((x-x_0)^2+(z-z_0)^2\right)^2}=\frac{2 m \sin 2\varphi}{l^2}, \label{eq:chainMF:Bx}
    \\
    B_z(x,y,z)&=-\frac{\partial\eta(x,y,z)}{\partial z}=
        \frac{2\, m\left((z-z_0)^2-(x-x_0)^2\right)}{\left((x-x_0)^2+(z-z_0)^2\right)^2}=-\frac{2 m \cos 2\varphi}{l^2} \label{eq:chainMF:Bz},\\
    \beta(\zeta)&=2\,m\,i \frac{(\zeta-\zeta_0)^2}{|\zeta-\zeta_0|^4}=\frac{2\,m\,i}{(\zeta^{*}-\zeta^{*}_0)^2},
    \label{eq:chainMF:zeta}
    \end{align}
here, $l$ and $\varphi$ define the cylindrical coordinate system
$(x-x_0)=l\cos\varphi,\quad (z-z_0)=l\sin\varphi$, and the angle
$\varphi$ is taken as counter-clockwise. Fig.\ref{Fig:duct:MF}a
shows an example of the magnetic field induced by two
symmetrically located chains of dipoles in the rectangular
channel.

 \begin{figure}
 \begin{center}
    \includegraphics[width=0.7\textwidth, angle=270]{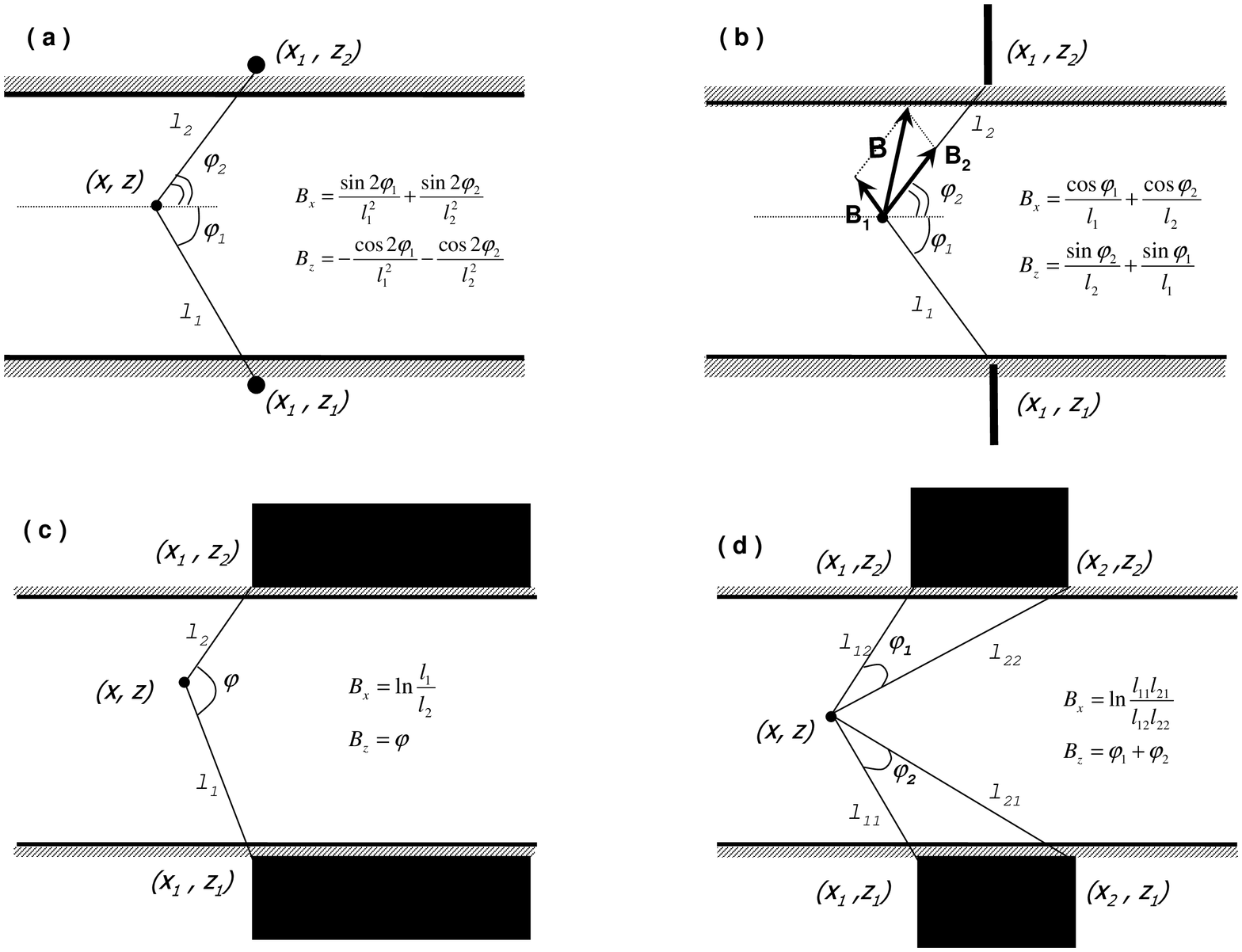}
    \caption{Inhomogeneous magnetic field created in the duct by
    linear chains of identical dipoles($a$), two identical sheets
    dipoles($b$), half space ($c$), and two magnets ($d$).
    Angles $\varphi$, $\varphi_1$, $\varphi_2$ must be taken as
    counter-clockwise, $m$ is taken equal one. Also, ($b$) shows
    a geometrical construction to find the final magnetic field,
    see Eq.~\ref{eq:sheetMF:Bx}-\ref{eq:sheetMF:Bz}} \label{Fig:duct:MF}
\end{center}
\end{figure}

The \textbf{sheet} of magnetic dipoles, Fig.\ref{Fig:duct:MF}$b$,
is made by stacking the linear chains described above along the
$z$ axis, $\Omega=\{x=x_0,-\infty\leq y \leq \infty, z_0 \leq z
\leq \infty\}$. Therefore for this case, the magnetic field is
obtained by an integration of
Eqs.~(\ref{eq:chainMF:eta}-\ref{eq:chainMF:zeta}) over $z$, from
$z_0$ to infinity:
    \begin{align}
    \eta(x,y,z)&= -m \ln \left( (x-x_0)^2+(z-z_0)^2\right)=2 m \ln \frac{1}{l}  \label{eq:sheetMF:eta}
    \\
    B_x(x,y,z)&=\frac{2 m(x-x_0)}{(x-x_0)^2+(z-z_0)^2}=\frac{2 m\cos\varphi}{l}, \label{eq:sheetMF:Bx}
    \\
    B_z(x,y,z) &= \frac{2 m(z-z_0)}{(x-x_0)^2+(z-z_0)^2}=\frac{2 m\sin\varphi}{l},\label{eq:sheetMF:Bz}\\
    \beta(\zeta)&= 2\,m\frac{\zeta-\zeta_0}{|\zeta-\zeta_0|^2}=\frac{2\,m}{(\zeta^{*}-\zeta^{*}_0)}.  \label{eq:sheetMF:zeta}
    \end{align}
This particular magnetic dipole  configuration allows the
construction of the final magnetic field through a clear
geometrical interpretation. Fig.~\ref{Fig:duct:MF}$b$ shows this
construction: the vector $\bf B_1$ is the field created by the
lower sheet and  is directed along the $\bf(r-r_1)$, while its
length is inverse to $l_1$; on the other hand, vector $\bf B_2$ is
the field created by the upper sheet, it points along the
$\bf(r-r_2)$ with a length inversely proportional to $l_2$; the
vector sum ${\bf B=B_1+B_2}$ is the final magnetic field in the
duct.

The \textbf{stepwise} magnetic field can be obtained by the
integration of the sheet fields from the edge point $x_1$ up to
infinity, ($\Omega=\{x_1\leq x \leq \infty, -\infty\leq y \leq
\infty, -\infty \leq z \leq z_1 \bigcup z_2 \leq z \leq \infty\})$,
see these coordinates in Fig.~\ref{Fig:duct:MF}c). To exclude the
divergence for $B_x$ at $x \rightarrow \infty$, we add a second
magnetic half-space, then:
    \begin{align}
    B_x(x,y,z)&= m\,\ln\frac{(x-x_1)^2+(z-z_1)^2}{(x-x_1)^2+(z-z_2)^2}=2\,m \ln\frac{l_1}{l_2}
    \label{eq:stepMF:Bx}
    \\
    B_z(x,y,z)&=  2\,m \left(\arctan\frac{z-z_1}{x-x_1}+\arctan\frac{z-z_2}{x-x_1}\right)
    =2\,m\, \varphi,    \label{eq:stepMF:Bz}\\
    \beta(\zeta)&= 2\,m\ln\frac{\zeta-\zeta_1}{\zeta-\zeta_2}    \label{eq:stepMF:zeta}.
    \end{align}
This, so called fringing, magnetic field is a candidate
configuration for the proper study of the transformation of the
Poiseuille velocity profile to the Hartmann one, and the
dependence of a transition length as a function of $Re$ and $Ha$.
To our knowledge there are no papers to date that apply a
physically correct fringing magnetic field in numerical
simulations, even though this field is one of the most heavily
studied in liquid metal flows, see
\cite{Molokov:Reed:Fusion:2003}, \cite{Alboussiere:2004},
\cite{Kumamaru:etal:2004}, \cite{Kumamaru:etal:2007},
\cite{ni_current_2007}. Below we derive qualitatively what is the
effect of the streamwise magnetic component of the fringing field
on the transverse pressure distribution in the duct.

The \textbf{shelfwise} magnetic field, created by a homogenous
dipole distribution, is obtained if  $\Omega$ is restricted in the $x$
direction ($\Omega~=~\{x_1 \leq x \leq x_2,-\infty\leq y \leq
\infty,
 -\infty\leq z \leq z_1 \cup z_2 \leq z \leq \infty\}$).
Fig.~\ref{Fig:duct:MF}d shows this configuration:
    \begin{align}
        \begin{split}
        B_x(x,y,z)&=\frac{m}{2}\ln\frac{l_{11} l_{21} }{l_{12} l_{22}}
         \\
            &= \frac{m}{2}\ln\frac{((x-x_1)^2+(z-z_1)^2)((x-x_2)^2+(z-z_1)^2)}
            {((x-x_1)^2+(z-z_2)^2)((x-x_2)^2+(z+z_2)^2)}    \label{eq:shelfMF:Bx}
        \end{split}
        \\
        \begin{split}
        B_z(x,y,z)&= m(\varphi_1+\varphi_2)\\
                  &= m \left( \arctan\frac{z-z_2}{x-x_1}-\arctan \frac{z-z_2}{x-x_2}
        + \arctan \frac{z_1-z}{x-x_1} - \arctan \frac{z_1-z}{x-x_2}\right) \label{eq:shelfMF:Bz},
        \end{split}
        \\
        \beta(\zeta)&= m\ln\frac{(\zeta-\zeta_{11})(\zeta-\zeta_{21})}{(\zeta-\zeta_{12})(\zeta-\zeta_{22})} \label{eq:shelfMF:zeta}.
    \end{align}
By varying the parameters $x_1, x_2, z_1, z_2$, one can regulate
the width of the central region and the upward (outward) gradient
of the magnetic field.

In above examples, the $z$ integration is taken over two half-open
regions $\{-\infty \leq z \leq z_1 \bigcup z_2 \leq z \leq \infty
\}$. It is easy to perform the integration up to a finite z-value
instead of infinity, however, this case is not considered in
order to simplify formulae and keep the essence of the method.
Moreover, in practice, a magnetic system is supported by the yoke
made of soft iron. This effectively means a closure for the lines
of the magnetic field coming from the external side of the magnet
(where the internal side is adjoined to the duct), therefore, the
upper limit of the integration can be safely set as infinity.

Up to now, we have learned how to define an external magnetic
field by means of magnetic dipoles. Usage of magnetic dipole
language dictates the following scaling behavior for the magnetic
field far from the central point: $l^{-3}$ for a single dipole,
$l^{-2}$ for a linear chain, $l^{-1}$ for a sheet, and $\ln l$ for
a massive body, where $l$ is a distance from the object, (see
Fig.~\ref{Fig:duct:MF}). It's possible to make this slope even
more smooth, i.e. $l\ln l$, if, instead of the magnetic domain, we
take a cross section of constant current density $j$, taken
everywhere equal to one.  Physically, it means that instead of
using the permanent magnet considered before, we shall deal with a
conducting cable carrying the electric current of homogenous
density. The easiest way to understand how we could develop
mathematically the cable from the magnet is described next.

\begin{figure}
    \begin{center}
    \includegraphics[width=0.75\textwidth, angle=-90]{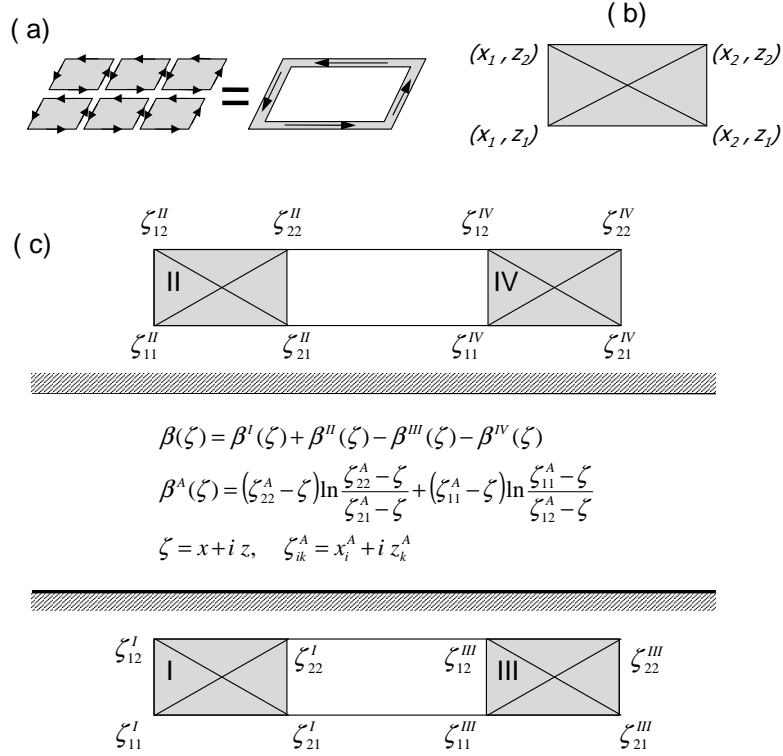}
    \caption{ Scheme showing an equivalence of the magnetic domain and a
    body with surface currents ($a$), notations used to write down magnetic field from
    a cable ($b$), scheme of two primitive solenoids located under and above duct($c$),
    the cables are supposed to encircle at $y=\pm\infty$, see text and
    Eq.~(\ref{eq:sol0MF:zeta},~\ref{eq:solMF:zeta})
    \label{Fig:domain2current}}
    \end{center}
    \end{figure}

First, we may present the magnetic domain as a non-conducting
cross-section enabling surface currents only. Schematically, it is
shown in Fig.~\ref{Fig:domain2current}$a$ by taking six magnetic
dipoles together in close contact: all the internal currents from
the adjoining magnetic dipoles are mutually compensated because
they run in opposite directions, therefore, the only non-compensated
electric currents are located on the bordering surfaces. The next step is
to extend the surface current obtained above, in such a way that
it flows through a finite rectangular region, i.e. through cable
of rectangular cross-section. The configuration of the rectangular
conductor is shown in Fig.~\ref{Fig:domain2current}$b$., where the
corners are $(x_k,z_l),\mbox{ } k,l=1,2$ (the diagonal cross in
Fig.~\ref{Fig:domain2current}$b$ denotes electric current running
in the direction perpendicular the plane of the figure). Such a
cable can be represented as a part of the top (cables II and IV)
and bottom (cables I and III) solenoids assembled around the
channel, see Fig.~\ref{Fig:domain2current}$c$.

To derive the magnetic field produced by the aforesaid primitive
solenoids, we note that the shelfwise magnetic system shown in
Fig.\ref{Fig:duct:MF}$d$,
Eqs.~(\ref{eq:shelfMF:Bx})-(\ref{eq:shelfMF:zeta}), is equivalent
to the solenoids having infinitely thin cables located at $x=x_1$
and $x=x_2$ because all internal currents in this system are
mutually compensated in the way explained above by
Fig.~\ref{Fig:domain2current}$a$. Therefore, if one integrates
$\beta(\zeta)$ given by Eq.~(\ref{eq:shelfMF:zeta}) over the
complex variable $\zeta=x+iz$ confined by cable's corners shown in
Fig.~\ref{Fig:domain2current}$b$, then one obtains the following
complex function for the individual cable $A$:
    \begin{eqnarray}
        \beta^A(\zeta)&=&(\zeta^A_{22}-\zeta)\ln\frac{\zeta^A_{22}-\zeta}{\zeta^A_{21}-\zeta}
                        +(\zeta^A_{11}-\zeta)\ln\frac{\zeta^A_{11}-\zeta}{\zeta^A_{12}-\zeta},
    \label{eq:sol0MF:zeta}
    \end{eqnarray} where $A=I,II,III,IV$ should be taken for different cables,
    and $\zeta^A_{k,l}=x_k^A + iz_l^A, \mbox{ } k,l=1,2$ are for
    the coordinates of the corners of cable $A$, see
    Fig.~\ref{Fig:domain2current}$c$. The final complex function
    for the top and bottom solenoids is obtained as a
    superposition of terms coming from four cables:
    \begin{eqnarray}
        \beta(\zeta)&=& \beta^{I}(\zeta) + \beta^{II}(\zeta) - \beta^{III}(\zeta) - \beta^{IV}
        (\zeta)\label{eq:solMF:zeta},
    \end{eqnarray}
where terms from cables I and II are positive and  for cables III
and IV are negative because the direction of the electric current
in cables I and II is opposite to the direction of current in
cables III and IV. The final $\beta(\zeta)$,
Eq.~\ref{eq:solMF:zeta}, is the linear superposition of the
products of logarithmic and rational functions depending on a
complex variable $\zeta=x+i\,z$. Then, by means of algebraic
transformations, the obtained $\beta(\zeta)$ should be presented
as a sum of its real and imaginary parts (\textit{cf.}
Eq.~\ref{eq:beta}), so that $B_x=\rm{Re}[\beta(\zeta)]$ and
$B_z=\rm{Im}[\beta(\zeta)]$. Although they have been obtained, the
mathematical expressions for $B_x$ and $B_z$ are quite cumbersome
and for the sake of brevity they are not reported here.

\begin{figure}
 \begin{center}
    \includegraphics[width=0.7\textwidth, angle=0]{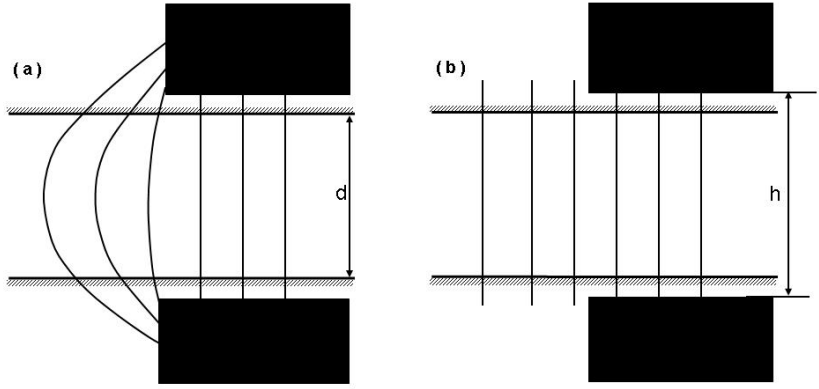}
    \caption{Pressure contour lines matching magnetic field
    lines with ($a$) and without ($b$) streamwise magnetic field
    component.
     \label{Fig:fringeMF}}
\end{center}
\end{figure}

To conclude, we show briefly what is the impact of the streamwise
field component on a liquid metal flow in the case of the fringing
magnetic field.  As it is widely accepted, inertia and viscosity
vanish in the core of a duct MHD flow, therefore, the pressure
distribution, $p$, is governed in the core by the Lorentz force,
$\nabla p={\bf j}\times{\bf B}$, where ${\bf j}$ are the induced
electric currents. This means that $\nabla p$ is perpendicular to
$\bf B$, i.e. the pressure contour lines  are matched with
magnetic field lines. Thus, if one employs the transverse field
component only, the pressure contour lines are straightened in the
transverse direction, while they must be curved in accordance with
the curvature of the external magnetic field, see
Fig.~\ref{Fig:fringeMF}.
Curvature of the magnetic field lines
depends on the aspect ratio $d/h$, where $d$ ($h$) is the height
of the duct (magnetic gap). As $d/h$ decreases, the curvature
vanishes, however  the inward gradient of the magnetic field
decreases as well. Hence, in order to study the effects of a
rapidly varying fringing magnetic field, the streamwise field
component must be taken into account necessarily.

\begin{figure}
 \begin{center}
    \includegraphics[width=0.7\textwidth, angle=0]{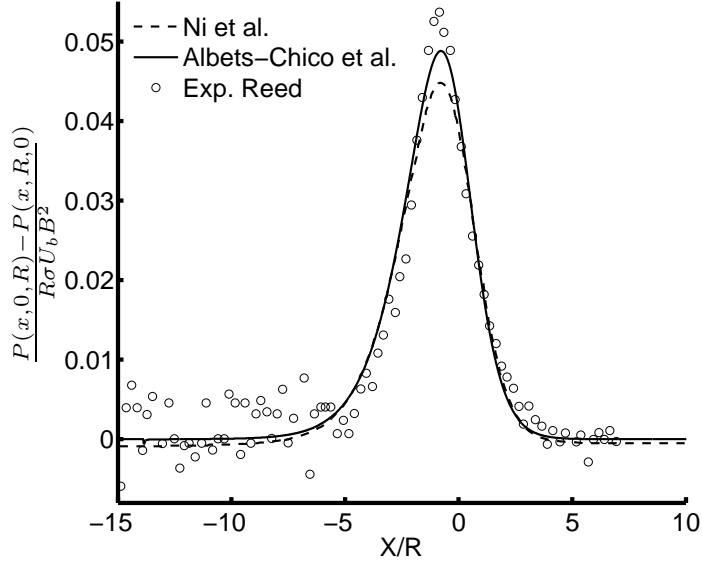}
    \caption{Dimensionless transverse pressure difference for the so called Benchmark
    fringing magnetic field case. Comparison is given for experimental
    results (\cite{Reed:Picologlou:etal:1987}, circles), 3D numerics with
    a non-consistent fringing magnetic field (dashed, \cite{ni_current_2007}), and
    3D numerics with a consistent magnetic field (solid, \cite{Albets-Chico:etal:2009})
    The results are clearly improved when the consistent magnetic field is considered.}
    \label{Fig:fringing_trans_pres}
\end{center}
\end{figure}

As a paradigmatic example, the well-known nuclear fusion benchmark
fringing magnetic field case based on the experiment of
\cite{Reed:Picologlou:etal:1987} ($Ha = 6569$, $N=10824$,
$c=0.027$) has been recently studied using  a complete numerical
resolution of the Navier-Stokes equations by
\cite{ni_current_2007}. Details regarding the shape of the
fringing magnetic field, geometry and boundary conditions for this
case are clearly exposed in both the experimental
\cite{Reed:Picologlou:etal:1987} and the numerical
\cite{ni_current_2007} works.

\cite{ni_current_2007} presented 3D numerical results in very good
agreement with the experimental data although a non negligible
under-prediction ($\approx 16\%$) of the peak transverse pressure
difference was also reported. Ni {\it et al.} applied a tanh-based
fitting function to approximate the main component of the
experimental magnetic field, while the additional components were
neglected. More recently, \cite{Albets-Chico:etal:2009} have also
addressed this case by means of a complete numerical resolution of
the governing equations (considering a quasi-static approximation
for the induced magnetic field) when using a nodal-based
non-structured finite-volume code. Details regarding the form of
the addressed Navier-Stokes equations and the assumed
simplifications can be found in  \cite{ni_current_2007}, as both
works have employed essentially the same methodology.
Additionally, Albets-Chico {\it et al.} have analyzed the effect
of the consistency of the magnetic field while developing a
mathematical technique to generate consistent magnetic fields from
experimental fits. Further, details will be available soon in
\cite{Albets-Chico:etal:2009}.

Fig. \ref{Fig:fringing_trans_pres} presents the effect of the
consistency of the magnetic field in the transverse pressure
difference results. When using a consistent magnetic field, the
under-prediction for the peak is reduced to approximately 9\%,
which clearly demonstrates how the consistency of the magnetic
field plays an important role related to the transverse pressure
difference in such a case. It is interesting to note that
Albets-Chico {\it et al.} obtained a perfect agreement with
\cite{ni_current_2007} results when using a non-consistent
magnetic field (based on the same tanh-based fitting function as
the one used in Ni {\it et al.}). Finally, Albets-Chico {\it et
al.} explain the still remaining under-prediction in terms of the
experimental results in the nature (slope and order of accuracy)
of the fitting function.

\begin{acknowledgement}
This work has been performed under the UCY-CompSci project, a
Marie Curie Transfer of Knowledge (TOK-DEV) grant (contract No.
MTKD-CT-2004-014199) funded by the CEC under the 6th Framework
Program. Partial support through a Center of Excellence grant from
the Norwegian Research Council to the Center for Biomedical
Computing is also greatly acknowledged. E.V.V. is grateful for
many fruitful discussions with Oleg Andreev, Yuri Kolesnikov,
Andre Thess, and Egbert Zienicke during his time in the Ilmenau
University of Technology.

\end{acknowledgement}

\bibliography{./../Bibtex/mhd}
\bibliographystyle{jfm}

\end{document}